\documentstyle[preprint,tighten,version2,aps]{revtex}
\def\goto{\mathop{\;\longrightarrow\;}}
{\makeatletter%
\gdef\labeleqs#1{{%
\edef\@currentlabel{%
\ifappendixon\appletter\fi
\ifsecnumbers\ifnum\c@secnum>0
\arabic{secnum}.\fi\fi\arabic{equation}}%
\label{#1}%
}}%
}
\begin{document}
\draft
\preprint{IFUP-TH 59/92}
\begin{title}
Finite size scaling in ${\rm CP}^{N-1}$ models
\end{title}
\author{Paolo Rossi and Ettore Vicari}
\begin{instit}
Dipartimento di Fisica dell'Universit\`a and
Istituto Nazionale di Fisica Nucleare, I-56126 Pisa, Italy
\end{instit}
\begin{abstract}
Finite size effects in Euclidean ${\rm CP}^{N-1}$ models with
periodic boundary conditions are investigated by means of the $1/N$
expansion and by Monte Carlo simulations. Analytic and numerical results
for magnetic susceptibility and correlation length
are compared and found to agree for small volumes.
For large volumes a discrepancy is found and explained as an effect of the
physical bound state extension.

The leading order finite size effects on the Abelian string tension are
computed and compared with simulations finding agreement. Finite size
dependence of topological quantities is also discussed.

\end{abstract}


\narrowtext
\section{Introduction}
\label{Introduction}
The study of the finite size effects has become an important
tool in the theoretical and numerical analysis of critical systems and
lattice field theories.

Finite size scaling may be used in order to improve the extrapolation of
finite lattice results to the infinite volume limit relevant to
field-theoretical
predictions. More naively, knowledge of the volume dependence of lattice
results in sufficiently large volumes allows to choose lattices such that
finite size effects may be consistently smaller than the intrinsical error
of numerical simulations.

Within our program of studying analytically and numerically
two-dimensional ${\rm CP}^{N-1}$ models as a prototype for lattice QCD,
we decided to explore their finite-volume behavior
with special emphasis on
two peculiar properties that our model share with QCD: existence of a
linear confining potential and non-trivial topological properties.

${\rm CP}^{N-1}$ models are $1/N$ expandable: we can therefore, for
sufficiently large $N$, make theoretical predictions on the size and
pattern of finite lattice effects. This chance, in conjunction with the
opportunity of performing high-statistics Monte Carlo simulations for
intermediate values of $N$ with an acceptable computer effort, offers the
possibility of a systematic study of finite size effects.

This papers is organized as follows:

In Sec. \ref{expansion1/N} we review the theory of finite size scaling in
the context of the $1/N$ expansion.

In Sec. \ref{Lattice} we discuss the lattice formulation adopted,
introduce the physical observables (magnetic susceptibility and
correlation length) and evaluate the $O(1)$ and $O(1/N)$ contributions
to their finite size functions.

In Sec. \ref{sigma} we perform the same analysis for the leading
order contribution to the Abelian string tension.

In Sec. \ref{simulations} we present the results of numerical
Monte Carlo simulations for selected values of $N$, from $N=2$ to
$N=100$, and compare them to our theoretical predictions, explaining
the discrepancies, when present.

In Sec. \ref{Conclusions} we draw some (hopefully general)
conclusions.

\section{Finite size scaling and $1/N$ expansion}
\label{expansion1/N}

Finite size scaling in the $1/N$ expansion has been introduced and
discussed in Ref.\cite{ONFSS}. We recall here the main results adapting them
to the specific context of ${\rm CP}^{N-1}$ models.
Any coordinate-independent physical quantity will in general depend on four
different parameters:
\begin{equation}
Q = Q(g,a,L,N)
\label{Qparam}
\end{equation}
where $g$ is the coupling, $L^d$ is the physical volume in $d$
dimensions and $a$ the lattice spacing. However in the critical region
$g\rightarrow g_c$ and in the infinite volume limit all dependence on $a$
and $g$ may be absorbed in the dependence on the correlation length
\begin{equation}
\xi \propto a\exp \int^g {dg^\prime\over \beta (g^\prime)}\;\;.
\label{xiscaling}
\end{equation}
The finite size scaling relation amounts to the statement that, when
$\xi\rightarrow \infty$ while keeping $L/\xi$ finite, we should get
\begin{equation}
{Q(g,a,L,N)\over Q(g,a,\infty,N)}\;\longrightarrow\;
f^{(Q)} (L/\xi,N)\;\;.
\label{FSSscaling}
\end{equation}
The $1/N$ expandability implies that, assuming
\begin{equation}
Q(g,a,L,N)= Q_0(Ng,a,L) + {1\over N}Q_1(Ng,a,L) +
O\left({1\over N^2}\right)\;\;
\label{lNexp}
\end{equation}
and
\begin{equation}
\xi(g,N)= \xi_0(Ng) + {1\over N}\xi_1(Ng) +
O\left({1\over N^2}\right)\;\;,
\label{lNexpxi}
\end{equation}
we may expand the finite size functions $f^{(Q)}$ in the form:
\begin{equation}
f^{(Q)}(L/\xi,N)=
f^{(Q)}_0(L/\xi) + {1\over N}f^{(Q)}_1(L/\xi)
+ O\left({1\over N^2}\right)\;\;,
\label{fexp}
\end{equation}
and obtain the representations:
\begin{equation}
{Q_0(Ng,a,L)\over Q_0(Ng,a,\infty)}\;\longrightarrow\;
f^{(Q)}_0(L/\xi_0)
\label{zeroorder}
\end{equation}
\begin{equation}
{Q_1(Ng,a,L)\over Q_0(Ng,a,L)} - {Q_1(Ng,a,\infty)\over Q_0(Ng,a,\infty)}
+ {L\over\xi_0}{f_0^{\prime {(Q)}}(L/\xi_0)
\over f_0^{(Q)}(L/\xi_0)}{\xi_1\over \xi_0}
\;\longrightarrow\; {f^{(Q)}_1(L/\xi_0)\over f^{(Q)}_0(L/\xi_0)}\;\;.
\label{firstorder}
\end{equation}

\section{Lattice formulation}
\label{Lattice}

Our choice of a lattice action for ${\rm CP}^{N-1}$ models is\cite{CPNlatt}:
\begin{equation}
S_{\rm g} = -{1\over g}\sum_{n,\mu}\left(
   \bar z_{n+\mu}z_n\lambda_{n,\mu} +
   \bar z_nz_{n+\mu}\bar\lambda_{n,\mu} - 2\right),\;\;\;\;\;\;
    g={1\over N\beta}\;\;,
\label{basic}
\end{equation}
where $z_n$ is an $N$-component complex scalar field, constrained by
the condition $\bar z_nz_n = 1$
and $\lambda_{n,\mu}$ is a ${\rm U}(1)$ gauge field satisfying
$\bar\lambda_{n,\mu}\lambda_{n,\mu} = 1\,$.

Observable physical quantities must be gauge invariant; we therefore focus
on the correlations of the composite operator
\begin{equation}
P_{i,j}(x) = \bar z_i(x) z_j(x)\;\;\;,
\label{projector}
\end{equation}
and specifically on the two-point function
\begin{equation}
G_P(x) = \langle {\rm Tr} P(x)P(0)\rangle _{c}\;\;.
\end{equation}
Starting from the continuum definition of magnetic susceptibility
\begin{equation}
\chi_m=\int d^2x \;G_P(x)
\end{equation}
and ($1/N$ expandable) correlation length \cite{CPN}
\begin{equation}
\xi^2={\int d^2x \;{1\over 4}x^2 G_P(x)\over \int d^2x\;G_P(x)}\;\;,
\end{equation}
we can introduce their counterparts on finite lattices with periodic
boundary conditions.
Let us first define the finite lattice Fourier transform of the correlation
function
\begin{equation}
\widetilde G_P(k;L) = {1\over L^2}\sum_{m,n} \langle {\rm Tr} P(m)
P(n)\rangle_{c} \exp \left[{2i\pi\over L}(m-n)\cdot k\right]
\end{equation}
defined on integer values of the components of $k$.
We then define
\begin{equation}
\chi_m(L) = \widetilde G_P(0,0;L)\goto_{L\rightarrow \infty} \chi_m\;\;\;,
\label{chim}
\end{equation}
\begin{equation}
\xi^2(L) = {1\over4\sin^2\pi/L} \,
\left[{\widetilde G_P(0,0;L)\over\widetilde G_P(1,0;L)} - 1\right]
\goto_{L\rightarrow \infty} \xi^2\;\;.
\label{xiG}
\end{equation}

Within the $1/N$ expansion the following relationship holds\cite{CPN}:
\begin{equation}
\widetilde G_P(k;L)\;=\;\beta^{-2}\;\left[
\Delta^{-1}_{(\alpha )}(k;L) + {1\over N}\Delta^{-1}_1(k;L)
+ O\left({1\over N^2}\right)\right]
\label{propN}
\end{equation}
where, in the infinite lattice notations:
\begin{equation}
\Delta^{-1}_{(\alpha)}(k)\;=\;\int {d^2 q\over (2\pi)^2}
{1\over \left[\widehat{q}^2+m_0^2\right]
\left[ \widehat{(k+q)}^2+m_0^2 \right]}\;\;\;,
\label{deltaalpha}
\end{equation}
\begin{eqnarray}
\Delta_1^{-1}(k)\;&&=\;
- \int {d^2 q\over (2\pi)^2} \Delta_{(\alpha)}(q) V_{(\alpha)}(q,k)
+ \int {d^2 q\over (2\pi)^2} \Delta_{(\lambda)}(q) V_{(\lambda)}(q,k)
\nonumber \\
+&& {1\over 2}\Delta_{(\alpha)}(0)
{\partial \over \partial m_0^2}\Delta_{(\alpha)}^{-1}(k)
\int {d^2 q\over (2\pi)^2} \left[
\Delta_{(\alpha)}(q)
{\partial \over \partial m_0^2}\Delta_{(\alpha)}^{-1}(q)
+\Delta_{(\lambda)}(q)
{\partial \over \partial m_0^2}\Delta_{(\lambda)}^{-1}(q)
\right]\;\;\;.
\label{Delta1}
\end{eqnarray}
The auxiliary quantities we have introduced are the effective vertices:
\begin{equation}
\Delta_{(\lambda)}^{-1}(k)\;=\; \int {d^2q\over (2\pi)^2}
{2\sum_\mu \cos q_\mu\over \widehat{q}^2+m_0^2}
-\int {d^2q\over (2\pi)^2} {4\sum_\mu \sin^2\left(q_\mu + k_\mu/2\right)
\over \left[\widehat{q}^2+m_0^2\right]
\left[\widehat{(q+k)}^2+m_0^2\right] }\;\;\;,
\end{equation}
\begin{eqnarray}
V_{(\alpha)}(k,p)\;&&=\; \int {d^2q\over (2\pi)^2}
{1\over \left[\widehat{q}+m_0^2\right]^2
\left[\widehat{(q+p)}^2+m_0^2\right]\left[\widehat{(q+k)}^2+m_0^2\right]}
\;+\;\; p\;\rightarrow -p \nonumber \\
+&& \int {d^2q\over (2\pi)^2}
{1\over \left[\widehat{q}+m_0^2\right]
\left[\widehat{(q+p)}^2+m_0^2\right]\left[\widehat{(q+k)}^2+m_0^2\right]
\left[\widehat{(q+p+k)}^2+m_0^2\right]}\;\;\;,
\end{eqnarray}
\begin{eqnarray}
V_{(\lambda)}(k,p)&=& \int {d^2q\over (2\pi)^2}
{4\sum_\mu \sin^2(q_\mu+p_\mu/2)
\over \left[\widehat{q}+m_0^2\right]^2
\left[\widehat{(q+p)}^2+m_0^2\right]\left[\widehat{(q+k)}^2+m_0^2\right]}
\;+\;\; p\;\rightarrow -p \nonumber \\
&+& \int {d^2q\over (2\pi)^2}
{4\sum_\mu \sin (q_\mu+p_\mu/2)\sin (q_\mu+k_\mu+p_\mu/2)
\over \left[\widehat{q}+m_0^2\right]
\left[\widehat{(q+p)}^2+m_0^2\right]\left[\widehat{(q+k)}^2+m_0^2\right]
\left[\widehat{(q+p+k)}^2+m_0^2\right]}\nonumber \\
&-&\int {d^2q\over (2\pi)^2} {2\sum_\mu \cos q_\mu
\over \left[\widehat{q}+m_0^2\right]^2
\left[\widehat{(q+k)}^2+m_0^2\right]^2}\;\;.
\end{eqnarray}
The conversion to finite lattice is obtained by the replacements
\begin{equation}
\int {d^2q\over (2\pi)^2} \longrightarrow
{1\over L^2}\sum_{q_\mu=1}^L\;\;,
\end{equation}
and on periodic (toroidal) lattices the zero modes of the
inverse propagator $\Delta_{(\lambda)}^{-1}(k)$,
if nonvanishing,  must be explictly removed.
Moreover
\begin{equation}
m_0^2\;\longrightarrow m_L^2\;\;,
\end{equation}
where in order to keep $Ng$ fixed we must relate $m_L$ to $m_0$ by the
lattice gap equation
\begin{equation}
{1\over L^2}\sum_{q_\mu=1}^L {1\over \widehat{q}^2+m_L^2}
\;=\;{1\over Ng}\;=\;
\int {d^2q\over (2\pi)^2} {1\over \widehat{q}^2+m_0^2}\;
=\;{1\over 2\pi}{1\over 1 + m_0^2/4 } K\left(
{1\over 1 + m_0^2/4 } \right)\;\;\;.
\label{gapequation}
\end{equation}
This equation has been analyzed in detail in Ref. \cite{ONFSS}
for $m_0\rightarrow 0$, i.e. in the scaling region, where
\begin{equation}
\int {d^2q\over (2\pi)^2} {1\over \widehat{q}^2+m_0^2}
\goto_{m_0\rightarrow 0} {1\over 4\pi}\ln {32\over m_0^2}\;\;\;.
\end{equation}
In the scaling region one could establish the relationship
\begin{equation}
z_0\;\equiv\; m_0L\;=\;z_c\exp -{\omega(z_L)\over 2}
\end{equation}
where $z_L=m_LL$, $z_c=4.163948...$, and
in the region $z_L \le 2\pi$ the function $\omega$ may be defined by
\begin{eqnarray}
\omega(z_L)&=& {4\pi\over z_L^2} + 4\pi\sum_{n=1}^\infty
(-1)^nz_L^{2n}d_{n+1}\;\;,  \nonumber \\
d_n &=& {1\over (2\pi)^{2n}}\sum_{m_1,m_2=-\infty}^\infty
{1\over (m_1^2 + m_2^2)^n}\;\;,\;\;\;\;\;n>1\;\;.
\end{eqnarray}
In the definition of $d_n$ the sum must be performed excluding the term
$(m_1,m_2)=(0,0)$.
The function $z_0(z_L)$ is monotonic and invertible. We could therefore
make use of the auxiliary variable $z_L$ in the computations
and replace it by $z_0$ only at the end.

In practice, in order to take into account at least some of the scaling
violation effects due to irrelevant operators, we found it always
convenient to solve explicitly and numerically Eq.\ (\ref{gapequation})
on each finite $L\times L$ lattice for assigned value of $Ng$.

Let us derive the large-$N$ results:
\begin{eqnarray}
\beta^2\chi_{m,0}(L)&=&\Delta_{(\alpha)}^{-1}(0,0;L)\;=\;
{1\over L^2}\sum_{q_\mu=1}^L {1\over (\widehat{q}^2+m_L^2)^2}\;=\;
-\,{\partial \beta\over \partial m_L^2}\;\;\;,\nonumber \\
\beta^2\chi_{m,0}&=&
\int {d^2q\over (2\pi)^2} {1\over \left(\hat{q}^2+m_0^2\right)^2}
\goto_{m_0\rightarrow 0} {1\over 4\pi m_0^2}\;\;\;,
\end{eqnarray}
and
\begin{eqnarray}
\xi_{0}^2(L) &&=\; {1\over4\sin^2\pi/L} \,
\left[{\Delta_{(\alpha)}^{-1}(0,0;L)\over
\Delta_{(\alpha)}^{-1}(1,0;L)} - 1\right]\;\;\;,\nonumber \\
\xi_{0}^2 &&={\rm lim}_{L\rightarrow\infty} \;\xi_{0}^2(L)
\goto_{m_0\rightarrow 0} {1\over 6m_0^2}\;\;.
\label{xi00}
\end{eqnarray}
The corresponding large-$N$ finite size functions are
obtained by the relations
\begin{eqnarray}
{\chi_{m,0}(\beta,L)\over
\chi_{m,0}(\beta,\infty)}&\rightarrow & f^{(\chi)}_0(L/\xi_0)\;\;\;,
\nonumber \\
{\xi_{0}(\beta,L)\over \xi_{0}(\beta,\infty)} &
\rightarrow & f^{(\xi)}_0(L/\xi_0)\;\;\;,
\label{f0}
\end{eqnarray}
for $\xi_0\rightarrow \infty$ while keeping $L/\xi_0$ constant.

The infinite volume limit quantities can be obtained numerically by
performing the calculation on very large lattices and checking the
convergence of the results.
For large enough lattices an extrapolation based on the formula
\begin{equation}
Q(L)\longrightarrow Q(\infty) + {c\over L^2}\;\;\;\;\;\;\;\;
{\rm for}\;\;\;\;\;L\gg 1\;\;\;,
\label{volextrap}
\end{equation}
gives safe results even if it does not keep into account the existing
logarithmic corrections to the $L^{-2}$ term.
Furthermore, we must in principle worry about the effects due to
irrelevant operators that may spoil the universal behavior
by introducing a separate dependence on $\xi_{0}^{-2}$.
Working at very large values of $\xi_0$ takes partially into account this
problem. Further improvement may be obtained by considering different
values of $\xi_{0}$ and linearly extrapolating in $\xi_{0}^{-2}$.
For $\xi_{0}$ large enough, this extrapolation, albeit purely
phenomenological because all existing logarithmic dependence on $\xi_{0}$
is not included, is powerful enough to predict results that are stable
under iteration of the procedure (within the numerical precision required).

An accurate evaluation of the functions $f_0^{(\chi)}$ and
$f_0^{(\xi)}$ is presented in Figs. \ref{LNFSSchi}
and \ref{LNFSSxi} respectively. The behaviors
are quite similar and are characterized by the existence of two different
regimes, corresponding to different physical situations.
As long as $L/\xi_0$ is small ($L < 5\xi_0$ say ), we are exploring the small
volume behavior of the system. Because of asymptotic freedom,
this is an essentially perturbative, small coupling regime, and actually
the finite size functions might be quite accurately evaluated within
standard finite volume perturbation
theory\cite{Luscher,Hasenfratz,Flyvbjerg},
with the only limitation of
ignoring the absolute scale of the abscissa, since perturbation theory
does not allow for the computation of the ratio between the mass gap
and the $\Lambda$ parameter which relates the quantity $\Lambda L$
appearing in perturbation theory to the finite size variable $L/\xi_0$.
When $L/\xi_0$ becomes large ($L> 10\xi_0$ say ) we start to explore
large-distance effects, that are dominated by the physical mass poles.
Due to the two-particle nature of the operator (\ref{projector}),
the functional dependence is characterized by a decrease with growing
$L/\xi_0$ (two particles in a box effect) in contrast with the asymptotic
growth characteristic of the corresponding single-particle operators
(free-particle in a box effect).

In the scaling region we can write $f^{(\chi)}_0$
in terms of $\omega(z_L)$; indeed we have
\begin{equation}
f^{(\chi)}_0(L/\xi_0)\goto_{m_0\rightarrow 0}
{\partial m_0^2 \over \partial m_L^2}\;=\;-z_0^2{\partial \omega(z_L)
\over \partial z_L^2 }\;\;,
\end{equation}
which is useful for $z_L < 2\pi$.
A similar formula can be obtained for $f_0^{(\xi)}$.

The determination of the $1/N$ corrections to the finite-size
functions proceeds according to the general formalism \cite{ONFSS}
described in the previous section.
Let us consider as an example the function $f^{(\chi)}_1$; the determination of
$f_1^{(\xi)}$ involves exactly the same steps and the details will not be
reported here. We may deduce from Eqs.
(\ref{firstorder},\ref{chim},\ref{propN})
the relationship
\begin{equation}
{f_1^{(\chi)} (L/\xi_{0})\over f_0^{(\chi)} (L/\xi_{0}) }\;=
\; {\Delta^{-1}_1(0;L)\over \Delta_{(\alpha)}^{-1}(0;L) }
- {\Delta^{-1}_1(0;\infty)\over \Delta_{(\alpha)}^{-1}(0,\infty) }
+ {L\over \xi_{0}}
{f_0^{\prime (\chi)} (L/\xi_{0})\over f_0^{(\chi)} (L/\xi_{0}) }
{\xi_1\over \xi_{0} }\;\;\;,
\end{equation}
where one can prove that
\begin{eqnarray}
\Delta_1^{-1}(0)\;=-{1\over 2}\int {d^2q\over (2\pi)^2} &&
\Delta_{(\alpha)}^{-1}(0) \Delta_{(\alpha)}(q)
{\partial\over \partial m_0^2}\left( \Delta_{(\alpha)}(0)
{\partial \over \partial m_0^2}\Delta_{(\alpha)}^{-1}(q) \right)
\nonumber \\
&& + \Delta_{(\alpha)}^{-1}(0) \Delta_{(\lambda)}(q)
{\partial\over \partial m_0^2}\left( \Delta_{(\alpha)}(0)
{\partial \over \partial m_0^2}\Delta_{(\lambda)}^{-1}(q) \right)
\;\;,
\end{eqnarray}
and its finite lattice generalization holds.
The function
\begin{equation}
H^{(\chi)}(L/\xi_{0})\;=\;
{1\over 2}{L\over \xi_0}
{f_0^{\prime (\chi)} (L/\xi_{0})\over f_0^{(\chi)} (L/\xi_{0}) }
\;=\;z_0^2 {\partial \over \partial z_0^2} \ln f_0^{(\chi)}
\end{equation}
could be extracted from the definition of $f_0^{(\chi)} (L/\xi_{0})$.
In particular one can show that in the scaling region
\begin{equation}
H^{(\chi)}(L/\xi_{0})\goto_{m_0\rightarrow 0}
1 - {L^2\over 2\pi}{\sum (q^2+m_L^2)^{-3}\over
\left[ \sum (q^2+m_L^2)^{-2} \right]^2 }\;=\;
1-{ \partial^2 \omega /(\partial z_L^2)^2 \over
\left( \partial \omega / \partial z_L^2 \right)^2 }\;\;.
\label{Hscaling}
\end{equation}
Finally we can evaluate
\begin{equation}
2\xi_0\xi_1\;=\;{\rm lim}_{L\rightarrow\infty}\;
{\Delta_{(\alpha)}^{-1}(0,0;L)\over \Delta_{(\alpha)}^{-1}(1,0;L) }
{1\over 4\sin^2 \pi/L } \left[
{\Delta_1^{-1}(0,0;L)\over \Delta_{(\alpha)}^{-1}(0,0;L) }-
{\Delta_1^{-1}(1,0;L)\over \Delta_{(\alpha)}^{-1}(1,0;L) } \right]\;\;.
\label{xi11}
\end{equation}
We are now in principle ready to evaluate the $1/N$ contribution to the
finite size functions of the magnetic susceptibility and of the correlation
length.

There is a further subtlety one must take into account when evaluating
quantities in a finite lattice.
The definition of $H^{(\chi)}$
implies that this quantity must vanish in the
$L/\xi \rightarrow \infty$ limit.
Contributions from irrelevant operators may therefore become comparable
to the value of the function itself in the large $L$ regime. We must
make sure that these contributions do not spoil
the final result. A very simple trick amounts to replacing the definition
extracted from Eq.\ (\ref{Hscaling}) with the following
\begin{equation}
H^{(\chi)}_L(z_0)\;=\;
1 - \left\{ { {1\over L^2} \sum (\widehat{q}^2+m_L^2)^{-3}\over
\left[ {1\over L^2} \sum (\widehat{q}^2+m_L^2)^{-2} \right]^2 } \right\}
\left\{ {\int  (\widehat{q}^2+m_0^2)^{-3}\over
\left[ \int (\widehat{q}^2+m_0^2)^{-2} \right]^2 } \right\}^{-1}\;\;.
\end{equation}
Since in the scaling region
\begin{equation}
{\int  (\widehat{q}^2+m_0^2)^{-3}\over
\left[ \int (\widehat{q}^2+m_0^2)^{-2} \right]^2 } \goto_{m_0\rightarrow 0}
2\pi\;\;,
\end{equation}
the two definitions differ only by irrelevant contributions.
However the second definition has the property that
it goes to zero when $L\rightarrow \infty$ even outside
the scaling region, and therefore it will not spoil the large $L$ behavior
of the function.

The same analysis goes through in the construction of the finite
size scaling function $f_1^{(\xi)}$ and we will not repeat it here.
Let us only quote the final expression we adopted in the numerical
calculations:
\begin{eqnarray}
&&H^{(\xi)}_L(z_0)\;=
z_0^2\partial \ln f_0^{(\xi)}/\partial z_0^2\;=
{1\over 2} + {1\over 4\sin^2 \pi/L} \;
{1\over 4\pi\xi_{0}^2(L) } \Delta_{(\alpha)}(1,0;L)\times \nonumber \\
&&\left[ \Delta_{(\alpha)}(1,0;L)
\,{1\over L^2}\sum {1\over \left[\widehat{q}^2+m_L^2\right]^2
\left[\widehat{(q+l)}^2+m_L^2 \right]} -
\Delta_{(\alpha)}(0,0;L)\,
{1\over L^2} \sum {1\over \left[\widehat{q}^2+m_L^2\right]^3 } \right]\;,
\end{eqnarray}
where $l=(1,0)$.

We are now ready for a numerical evaluation of $f^{(\chi)}_1/f^{(\chi)}_0$
and $f^{(\xi)}_1/f^{(\xi)}_0$. The results are presented in Figs.\
\ref{f1chi} and \ref{f1xi}
respectively. Values obtained at finite $\xi_0$, and therefore affected by the
presence of non-scaling contributions, are subsequently extrapolated at
infinite $\xi_0$, after having verified that scaling volations are well
parametrized by a term proportional to $\xi_0^{-2}$.
We estimate the error in our
determinations to be uniform and to be everywhere contained within the
graphical width of the line representing our results.

We notice that the behaviors are similar but they are not quite the same.
In particular $f^{(\chi)}_1$ has two zeros: the smaller one lies in the
perturbative region and should manifest itself in the existence of a region
where the finite size function at finite $N$ is, for $N$ large enough,
independent of $N$. This result will be confirmed by Monte Carlo
simulations in Sec. \ref{simulations}.

\section{String tension}
\label{sigma}

Another very important physical quantity in the context of
${\rm CP}^{N-1}$ models (and of lattice gauge theories) is the string
tension, that may be extracted from the asymptotic area law holding for
Wilson loops.
Within the $1/N$ expansion, the string tension is depressed by a power
of $N^{-1}$, and therefore it is sufficient for our purpose to focus
on the first non-trivial contribution to the rectangular Wilson loop:
\begin{equation}
\ln W(R,T)\;=\; - {1\over 2N} \int {d^2 k\over (2\pi)^2 }
{ \sin^2 {k_1R\over 2}\over \sin^2 {k_1\over 2}}
{ \sin^2 {k_2T\over 2}\over \sin^2 {k_2\over 2}}
\widehat{k}^2 \Delta_{(\lambda)}(k)\;\;\;.
\end{equation}
Actually in order to extract the string tension it is convenient to study
the Creutz ratios:
\begin{eqnarray}
\chi_C(R,T)&=& \ln {W(R,T-1)\,W(R-1,T)\over
W(R,T)\,W(R-1,T-1) }\nonumber \\
&=& {1\over 2N} \int {d^2 k\over (2\pi)^2 }
{ \sin {k_1\over 2} (2R-1)\over \sin {k_1\over 2}}
{ \sin {k_2\over 2} (2T-1)\over \sin {k_2\over 2}}
\widehat{k}^2 \Delta_{(\lambda)}(k)\;\;\;,
\label{Creutzratio}
\end{eqnarray}
and the Polyakov ratios
\begin{eqnarray}
\chi_P(R)&=& \ln {W(R-1,L)\over W(R,L)}\nonumber \\
&=&{1\over 2N} \int {d k_1\over 2\pi }
{ \sin {k_1\over 2} (2R-1)\over \sin {k_1\over 2}}
\widehat{k_1}^2 \Delta_{(\lambda)}(k_1,0)\;\;\;.
\label{Polyakovratio}
\end{eqnarray}
In the infinite volume limit and in the scaling region,
 for $R,T\gg \xi_{0}$ one obtains
\begin{eqnarray}
\chi_C(R,T)&\longrightarrow &
{6\pi m_0^2\over N} \;=\;{\pi\over N \xi_{0}^2 }\;\equiv\;\sigma_0
\nonumber \\
\chi_P(R) &\longrightarrow &\sigma_0\;\;\;.
\end{eqnarray}
On a finite lattice $\chi_C$ and $\chi_P$ are obtained by replacing
in Eqs.\ (\ref{Creutzratio},\ref{Polyakovratio})
$\Delta_{(\lambda)}(k)$
with the finite size propagator $\Delta_{(\lambda)}(k;L)$.
We may focus on the square Wilson loops ($R=T$)
and compute the dimensionless quantity $C(R)=\chi_C(R,R)\xi^2$.
However, when studying finite size effects, we must take into account the
existence of two different physical length scales: $\xi$ and $R$.
Therefore, even in the scaling region we end up in a function
of two variables; indeed for $\xi\rightarrow \infty$ keeping fixed
the ratios $L/\xi$ and $R/\xi$ we must have
\begin{equation}
C(N,g,R,L,a)\goto_{\xi\rightarrow\infty}f^{(C)}(N,L/\xi,R/L)
\goto_{N\rightarrow \infty} {1\over N}f^{(C)}_0(L/\xi_0,R/L)\;\;.
\end{equation}
The same relation holds for the dimensionless quantity
$P(R)=\chi_P(R)\xi^2$:
\begin{equation}
P(N,g,R,L,a)\goto_{\xi\rightarrow\infty}f^{(P)}(N,L/\xi,R/L)
\goto_{N\rightarrow \infty} {1\over N}f^{(P)}_0(L/\xi_0,R/L)\;\;.
\end{equation}

We evaluated the function $f^{(P)}_0$ numerically at the fixed value
$\beta=0.86$ such that $\xi_0\approx 16$, where we checked that scaling was
satisfied within $0.2\%$.
The results are plotted in Fig.\ \ref{fP0} for several values
of $L$ as a function of $R/L$. The resulting curves correspond to the
sections at constant $L/\xi_0$ of the two-variable function $f^{(P)}_0$.
This choice of variables is most adequate if we want to focus on large
Wilson loops (i.e. such that $R\gg\xi$). Otherwise a more proper choice
would have been the couple $L/\xi_0$, $R/\xi_0$.

Notice that even in the limit $L/\xi_0,R/\xi_0\rightarrow\infty$
as long as the ratio $R/L$ is nonvanishing, we do get a nontrivial function
of this ratio.
Since at large $N$, in the continuum and for infinite volume,
we have that for $k\xi_0\rightarrow 0$
\begin{equation}
\Delta_{(\lambda)}(k)\goto
{2\pi\over \xi_{0}^2 {k}^2 }\;\;,
\end{equation}
we may exploit the condition
$L/\xi_0\rightarrow\infty$ with $R/L$ finite
to replace the dressed propagator $\Delta_{(\lambda)}(k;L)$ with
a lattice representation of its pole part\cite{CPNlatt2}:
\begin{equation}
\Delta_{em}(k)\;=\;{2\pi\over \xi_{0}^2 \widehat{k}^2 }\;\;.
\end{equation}
It is then easy to show that
\begin{equation}
{\rm lim}_{L/\xi_{0}\rightarrow \infty} \;\;f_0^{(C)}(L/\xi_{0},R/L)
\;=\;\pi \left[ 1 - \left( {2R-1\over L} \right)^2 \right]\;\;,
\end{equation}
while the corresponding computation for the Polyakov loops leads to
\begin{equation}
{\rm lim}_{L/\xi_{0}\rightarrow \infty} \;\;f_0^{(P)}(L/\xi_0,R/L)
\;=\;\pi \left[ 1 - \left( {2R-1\over L} \right) \right]\;\;.
\end{equation}
Actually one may easily prove that the vanishing of the Creutz and Polyakov
ratios at $R=(L+1)/2$ holds for any finite lattice and for any acceptable
choice of the propagator $\Delta_{(\lambda)}$, and depends only on the
Abelian structure of the loop variables.

We may study the lattice symmetries and further notice that
\begin{equation}
f_0^{(P)}(R) \;=\;-f_0^{(P)}(L+1-R) \;\;\;,
\end{equation}
while, defining the combination
\begin{equation}
\chi_{sym}(R,T)\;=\;\chi_C(R,T)-\chi_P(R)-\chi_P(T)\;\;,
\end{equation}
 we find
\begin{equation}
\chi_{sym}(R,T)\;=\;\chi_{sym}(L+1-R,L+1-T)\;\;.
\end{equation}
Hence
\begin{equation}
f^{sym}_0(R)\;=\;f^{sym}_0(L+1-R)\goto_{L/\xi\rightarrow\infty}
-\pi \left[ 1 - \left( {2R-1\over L} \right) \right]^2\;\;.
\end{equation}
It is tempting to redefine the finite size scaling functions extracting the
purely kinematical factors derived above.
We shall consider for definiteness the quantity (even under the exchange
$R\rightarrow L+1-R$)
\begin{equation}
g_0^{(P)}(L/\xi_{0},R/L)\;=\;
{f_0^{(P)}(L/\xi_{0},R/L)\over 1 - \left( {2R-1\over L}\right)}
\goto_{L/\xi_{0}\rightarrow \infty} \pi\;\;\;.
\end{equation}
When $L\gg \xi_{0}$ we can study the regime $L > 2R\gg\xi_{0}$, expecting
our results to be largely independent of the ratio $R/\xi_{0}$.
Naively this might appear to imply that $g_0^{(P)}(L/\xi_{0},R/L)$
should not depend on $R/L$.
However this condition is realized in the following subtler way:
\begin{equation}
g_0^{(P)}(L/\xi_{0},R/L)
\goto
\pi + \phi(L/\xi_{0})\psi(R/L)
\end{equation}
for $L/\xi_{0}\gg 1$ and $R/\xi_{0}>1$,
and the function $\phi(L/\xi_{0})\rightarrow 0$ for
$L/\xi_0\rightarrow\infty$.

The function $g^{(P)}_0$ has been numerically obtained directly from the
abovementioned determination of $f^{(P)}_0$ and is plotted in Fig.\
\ref{gP0}.
We extracted our best determinations of the functions $\phi$ and $\psi$
from an analysis of the same numerical data, when they were relevant
(i.e. for $R/\xi_0 > 1$). Our results are presented in Figs. \ref{phi}
 and \ref{psi},
where we have chosen the normalization $\psi(1/2)=1$.
Studying the Creutz ratios we found a similar behavior.

The origin of the nontrivial albeit kinematical (i.e.
not related to the propagation of the physical degrees of freedom)
dependence in the regime $R/\xi_0>1$ can be traced qualitatively
to the observation that, for all finite lattices, the dynamically generated
massless pole in the propagator $\Delta_{(\lambda)}(k)$ does in fact
disappear, and the linear confining potential is replaced,
in the small $k$ regime, by an effective
Yukawa potential parametrized by a ``mass'' $\mu=\mu(L/\xi_0)$
becoming smaller and smaller with increasing $L$, and therefore
approximating the linear behavior with better and better accuracy
according to the relationship
\begin{equation}
{\rm lim}_{\mu\rightarrow 0}{1-e^{-\mu R}\over \mu}\;=\;R\;\;\;.
\end{equation}
The parameter $\mu(L/\xi_0)$ sets the scale of an effective
long range lattice interaction; when $L/\xi_0\rightarrow \infty\;\;$
$\mu(L/\xi_0)\rightarrow 0$, and the infinite volume physics is well
approximated when $1/\mu\gg L$;
but when $L/\xi_0$ is not too large the
relationship $L > 1/\mu \gg \xi_0$ may hold.
As a consequence of this phenomenon, the measured finite lattice string
tension, whose operational definition might be
\begin{equation}
g_0^{(P)}(L/\xi_{0},R/L=1/2)
\goto \pi + \phi(L/\xi_{0})\;\;\;,
\end{equation}
may turn out to be significantly different from $\pi$
even for large values of $L/\xi_{0}$.

In order to have a more quantitative perception of this phenomenon,
we performed a study of the finite size dependence of the function
$\Delta_{(\lambda)}^{-1}(k,0;L)$, plotted in Fig.\ \ref{deltak}
as a function of the dimensionless variable $k\xi_0$
for several values of $L/\xi_0$. Notice that the variable $k\xi_0$
can only take discrete values, and for small $L/\xi_0$ these values
are quite distant from each other.
In Fig.\ \ref{delta0}
we plot the values of $\Delta_{(\lambda)}^{-1}(0;L)\equiv
\mu^2\xi_0^2/2\pi$ against the function $(\xi_0/L)^2$ taken as an estimator
of the contribution from the first nontrivial value of $k$.
Fig. \ref{delta0} shows that the condition $\Delta_{(\lambda)}^{-1}(0;L)
\ll (\xi_0/L)^2$ (corresponding to $L\ll 1/\mu$) holds for
$L/\xi_0 \ge 25$, while for smaller $L/\xi_0$ significant deviations from the
infinite volume behavior are to be expected, as already phenomenologically
observed.

We evaluated analitically the function $g^{(P)}_Y(L/\xi_0,R/L)$
in the scaling limit, assuming an effective Yukawa propagator,
and found
\begin{equation}
g^{(P)}_Y(L/\xi_0,R/L)\;=\;\pi{{\rm sh} \left[ {\mu L \over 2}
\left(1-{2R\over L}\right) \right] \over {\rm sh} \left({\mu L\over 2}\right)
\left( 1-{2R\over L}\right) }\;\;\;,
\end{equation}
where
\begin{equation}
(\mu L)^2\;=\;2\pi\Delta_{(\lambda)}^{-1}(0;L/\xi_0) \left({L\over \xi_0}
\right)^2\;\;\;.
\end{equation}
The corresponding functions
\begin{equation}
\phi_Y(L/\xi_0)\;=\;\pi \left[ { {\mu L\over 2}
\over {\rm sh} ({\mu L\over 2}) }
-1\right]
\end{equation}
and $\psi_Y(R/L)$ are plotted in Figs. \ref{phiY} and \ref{psiY},
using the same conventions and notations of Figs. \ref{phi}, \ref{psi}.
The qualitative and quantitative agreement is extremely satisfactory,
in view of the crude approximations, and confirms our
interpretation of this finite size effect.

Up to now we have assumed to be in the scaling region and therefore we have
neglected the scaling deviations, which, at finite $\xi_0$,
generate corrections
to the value of the string tension.
By using the relation
\begin{equation}
{\chi_P(\beta,R,L)\over \left[ 1 -\left( {2R-1\over L}\right) \right] }
\goto \sigma_0(\beta)
\end{equation}
for $L,R\gg \xi_0$,
we extracted the value of the string tension for several values
of $\beta$. The results for the dimensionless quantity
$N\sigma_0\xi_0^2$ are reported in Table \ref{tab_sigma} and
plotted in Fig.\ \ref{sigma0} versus $1/\xi^2_0$.
They converge to the continuum value $\pi$, and,
as expected, the scaling violations are approximatively
proportional to $\xi^{-2}_0$.
{}From the square Creutz ratios (i.e. those with $R=T$)
we obtained the same results via the relation
\begin{equation}
{\chi_C(\beta,R,R,L)\over \left[ 1 -\left( {2R-1\over L}\right)^2 \right] }
\goto \sigma_0(\beta)
\label{sigmaCR}
\end{equation}
for $L,R\gg \xi_0$.

\section{Simulations}
\label{simulations}
In order to check the predictive power of the $1/N$ expansion approach
to finite-size scaling and to obtain results for small values of $N$
where the expansion is not expected to give sensible information, we
proceeded to Monte Carlo simulations of ${\rm CP}^{N-1}$ models
for a wide spectrum of values of $N$.
In the following we will show numerical results for $N=2,4,10,21,41,100$.
Data for $N=2$ and $N=4$ were already presented in Refs.
\cite{CPNlatt,CPNlatt2}. Summaries of the results for $N=10,21,41,100$
are given respectively in Tables \ref{tab_cp9}, \ref{tab_cp20},
\ref{tab_cp40}, \ref{tab_cp99}.

The action (\ref{basic}) is known to enjoy precocious scaling, and the
chosen values of $\xi$ are expected to be consistent with scaling
violations $O(1\%)$ for all the observables derived from the two point
function $G_P$ \cite{CPNlatt,CPNlatt2,CPNlatt3}.
In most of our
simulations we employed algorithms consisting in efficient mixtures
of over-heat bath and microcanonical algorithms. The detailed
description of this simulation algorithm with a discussion of its
dynamical features is contained in Ref. \cite{CPNlatt}.
Data relative to the largest lattices for $N=21$ and $N=41$ are taken
from Ref. \cite{CPNlatt3} where the simulated tempering algorithm
\cite{Parisi} was employed.

By varying the lattice size at fixed $\beta$ we managed to reconstruct
the finite size functions $f^{(\chi)}$ and $f^{(\xi)}$.
They were obtained by approximating infinite lattice quantities with
the corresponding values measured on the largest lattice
available for each $\beta$ and $N$ (of course after checking the
stability of the results on large lattices).
The results relative to $f^{(\chi)}$ and $f^{(\xi)}$ are plotted
respectively
in Figs. \ref{fchi100} and \ref{fxi100} for $N=100$,
in Figs. \ref{fchi41} and \ref{fxi41} for $N=41$,
in Figs. \ref{fchi21} and \ref{fxi21} for $N=21$,
in Figs. \ref{fchi10} and \ref{fxi10} for $N=10$.
In these figures the dashed lines show the large-$N$ predictions
(\ref{f0}), and the continuous lines represent the inclusion
of the $1/N$ correction evaluated in Sec. \ref{Lattice}.

In the scaling region the finite size scaling functions must be universal,
that is independent of $\beta$ and of the lattice formulation, in that
they should reproduce the continuum physics in a periodic box.
Data for $N=10,21,41$ coming from different $\beta$ values show
this universal behavior.

A summary of all the most significant results is presented in
Figs. \ref{fchi} and \ref{fxi} where all different values of $N$ are
brought together, including data for $N=2$ and $N=4$, in order
to see the general trends.

As anticipated in Sec. \ref{Lattice}, two physically distinct regions
exist.
In the small volume domain the $1/N$ expansion
(as well as the perturbative expansion) is predictive for values
of $N$ as small as 10. The vanishing of $f^{(\chi)}_1$ in
a neighbourhood of $L/\xi\simeq 3$ reflects itself
in the approximate universality of the corresponding finite size
curves in the same region for all $N\ge 4$.
The large volume behavior shows a peculiar disagreement with the $1/N$
expansion: convergence to the infinite volume prediction is much
faster than expected even for rather large values of $N$ and moreover,
it follows a pattern that does not seem to be amenable to analytic $1/N$
effects.

Actually there is a plausible interpretation of this phenomenon,
as we shall try to show. One must not forget that the physical spectrum
of ${\rm CP}^{N-1}$ models is characterized by effects of confinement,
that in turn leads to the appearance of two-particle bound
states, dominating the large distance behavior of the function $G_P$.
These bound states have binding energies order $(6\pi/N)^{2/3}$ that
are nonanalytic in $1/N$, and physical extension (effective radius)
order $(N/6\pi)^{1/3}$.
Therefore they will dominate the finite size effects on
large lattices ($L\sim N^{1/3}\xi$), while in the intermediate region the
finite size functions must interpolate between the analytic
(small volume) and the nonanalytic behavior.
We may therefore expect these functions, for intermediate values of $N$,
to show a (roughly) universal behavior for intermediate and large
values of the effective variable $L/(\xi N^{1/3})$. This effect is shown
in Figs. \ref{fchivrs} and \ref{fxivrs}, where this kind of universality
is shown to apply to values $4\le N \le 41$.
There is necessarily a saturation effect, because otherwise the convergence
to infinite volume results would be further delayed for larger values
of $N$, in contrast with the $N\rightarrow \infty$ prediction.
Indeed at $N=100$ we see a significant departure from the $\xi N^{1/3}$
scaling and an improved agreement with the $1/N$ expansion
even in the large volume regime.

At the opposite end of our analysis, when $N=2$  we find that none of the
previous considerations apply, but for the obvious agreement
at small $L/\xi$ due to the predictivity domain of standard perturbation
theory (that commutes with the $1/N$ expansion).
${\rm CP}^1$ is equivalent to $O(3)$ non linear $\sigma$ model,
and the finite size curves are actually showing the typical behavior of
single-particle correlations in a finite volume (free particle in a box).
We stress however that this different qualitative pattern seems to come out
without any quantitative discontinuity in the dependence on $N$, that
appears to be monotonic and smooth in the whole range of the variable.

The infinite lattice correlation
lengths obtained from our Monte Carlo simulations
can be compared with those predicted by the $1/N$ expansion.
{}From Eqs. (\ref{xi00},\ref{xi11}) we can get $\xi(\beta)$
up to $O(1/N)$. For $N=100$ and $\beta=0.54$ we find
$\xi=1.9106$, to be compared with the corresponding
Monte Carlo result on the largest available lattice: $\xi=1.922(9)$
(see Table \ref{tab_cp99}).
In the continuum the $1/N$ expansion of $\xi$ leads to \cite{CPN2}:
\begin{equation}
\xi\Lambda_{L} \;=\; {1\over 8 \sqrt{3}}\left[ 1 - {6.7033\over N}
+ O\left({1\over N^2}\right)\right]\;\;\;,
\end{equation}
where $\Lambda_{L}$
is the $\Lambda$ parameter of the lattice action (\ref{basic}).
Assuming asymptotic scaling, according to the two loop formula
\begin{equation}
g(\beta)\;=\;(2\pi\beta)^{-2/N}\exp (2\pi\beta )\;\;\;,
\end{equation}
we can find the correlation length as a function of $\beta$:
$\xi(\beta )=(\xi\Lambda_L) g(\beta )$.
For $N=100$ and $\beta=0.54$ this would give $\xi=1.955$,
consistent with $O(1/N\beta )$ deviations from asymptotic scaling.

The string tension is easily extracted by measuring the
square Creutz ratios $\chi_C(R,R)$; indeed for $L,R\gg \xi$
we should have
\begin{equation}
g^{(C)}(R)\;=\;
{\chi_C(R,R)\over \left[ 1 -\left( {2R-1\over L}\right)^2 \right] }
\xi^2\goto \sigma\xi^2\;\;\;.
\end{equation}
As shown in Sec. \ref{Lattice}, in practice the factor
$\left[ 1 -\left( {2R-1\over L}\right)^2 \right]$ allows us to perform the
measurement without requiring $L\gg R$.
The string tension has been already measured for $N=4$, $N=10$
\cite{CPNlatt2}, $N=21$, and $N=41$ \cite{CPNlatt3}, showing
for the last two values of $N$
agreement with the large-$N$ prediction $\sigma\xi^2=\pi/N$.
We have repeated the measurement for $N=100$,
at $\beta=0.54$ on a $L=48$ lattice,
corresponding to $L/\xi\simeq 25$.
In Fig. \ref{sigma100} $g^{(C)}$
is plotted versus the physical distance $r=R/\xi$;
the continuous and the dashed
lines show the large-$N$ predictions
obtained from Eq. (\ref{Creutzratio})
at the same bare coupling $\beta=0.54$,
respectively for an infinite and a $L=48$ lattice.
The analysis done in Sec. \ref{sigma} and the
small differences between the infinite and the $L=48$ lattice
predictions
show that $L/\xi\simeq 25$ should be sufficient
to determine the string tension through the
measure of $g^{(C)}$.
Monte Carlo data are consistent with
the large-$N$ results, which also predict the observed discrepancy
of the measured string tension
from the continuum value $\sigma\xi^2=\pi/N$ due to scaling violations
(see Table \ref{tab_sigma} and  Fig. \ref{sigma0}).

Finally we would like to make a few remarks on the study of the
topological properties of ${\rm CP}^{N-1}$ models on the lattice.
At large $N$ a particularly severe form of critical slowing down,
exponential with respect the correlation length, has been observed
when measuring the topological susceptibility \cite{CPNlatt}.
Moreover, this phenomenon becomes worse with increasing $N$,
making the simulations effectively non ergodic, already
at small $\xi$.
In order to overcome this difficulty and to perform simulations sampling
correctly the topological sectors at large $N$, in Ref. \cite{CPNlatt3}
the simulated tempering method of updating \cite{Parisi}
was employed, and agreement
with the large $N$ result $\chi_t\xi^2=1/(2\pi N)$ was found for $N=21$ and
$N=41$.

At large $N$ we expect that the dynamically generated gauge field
shows a phenomenon similar to that observed in the
two-dimensional $U(1)$ gauge model
\cite{Smit}, that is a strong decoupling in the infinite volume limit
between the gaussian modes around the large-$N$ saddle point, responsible
for confinement, and the modes determining the topological properties.
A manifestation of this decoupling would be the possibility
to obtain good estimates of observables not related to the topological
properties even in simulations not sampling correctly the topological
sectors \cite{CPNlatt3}. Relying on this decoupling property
we decided to use a local algorithm to perform our simulations
at large $N$.

For $N=100$ different topological
sectors are separated by such a high free energy barrier that starting
from a flat configuration, even after a long thermalization procedure,
we found zero geometrical charge for each configuration generated
in our simulations.
All our $N=100$ results reported in Table \ref{tab_cp99}
are effectively obtained by sampling configurations with $Q_g=0$.
We also found that the geometrical charge $Q_g$ remained unchanged even
when starting from a configuration having $Q_g\neq 0$. As expected,
the measured values of $\chi_m$, $\xi$ and $\chi_C(R,R)$ with $R\ll L$
become independent of
$Q_g$ for large enough lattices (and for reasonable
values of $Q_g$).
Anyway,
since the large-$N$ saddle point configuration has trivial
topology, for all finite lattices the correct procedure to obtain
reliable results is starting from a flat configuration and therefore
effectively sampling configurations with $Q_g=0$.
The good agreement of the numerical results obtained in this way
with the large-$N$ predictions
is a further support to the abovementioned decoupling phenomenon.

\section{Conclusions}
\label{Conclusions}

The analytical and numerical results we have presented lead
to some considerations whose relevance might go beyond the domain of
${\rm CP}^{N-1}$ models.

I) There is a perturbative domain for finite size effects, where
it is possible to compute the finite size functions with satisfactory
precision, both in the $1/N$ and in the standard weak coupling expansion.
For asymptotically free theories, this domain coincides with the small
volume regime. In the spirit of L\"uscher and collaborators'
investigations \cite{Luscher,Luscher2},
one might conceive of supplementing the perturbative
evaluations with numerical simulations at rather small volumes and
matching numerical finite size effects (functions of $L/\xi$)
with analytical results (functions of $\Lambda L$) in order to extract
physical information (e.g. the quantity $\Lambda\xi$).

II) In the presence of a confining potential,
bound states have  a physical radius that can be significantly bigger
than the correlation length appearing in the long distance behavior.
This new physical scale may dominate the finite size effects on volumes
that are large with respect to the correlation length but comparable with the
bound state radius itself.

III) Phenomena related to topological properties show peculiar finite
lattice behaviors.
Actually
our calculations have shown that assuming to define a topological
susceptibility by the standard (continuum or infinite lattice) procedure
\begin{equation}
\chi_t\;=\; {\rm lim}_{k\rightarrow
0}\;\;\widehat{k}^2\Delta_{(\lambda)}(k)\;\;\;,
\end{equation}
the result would be zero on any finite lattice because the dynamically
generated massless pole disappears on finite lattices, as expected
on general theoretical grounds. As we have seen, this is also the reason
why it is strictly speaking impossible to give a definition of the string
tension on finite lattices.
However one must not forget that the infinite volume limit is reached
smoothly and the convergence, for sufficiently large $L$, becomes
exponentially fast (water boils also in your pot, and not only in the
infinite volume limit!). Therefore it should always be possible, with an
appropriate limiting procedure, to extract infinite volume information
from finite volume simulations. We have seen that careful handling
leads to the possibility of ``measuring'' the string tension on finite
lattices within the desired precision.
Similarly, we see no obstruction to interpreting the so-called
``geometrical'' definitions of topological charge (and derived quantities)
as limits of sequences of local operators \cite{CPNlatt}
that can be approximately evaluated on finite lattices, thus giving
a precise meaning to apparently ill-defined quantities.



\figure{Large-$N$ finite size scaling function
of the magnetic susceptibility.
\label{LNFSSchi}}

\figure{Large-$N$ finite size scaling function of the correlation length.
\label{LNFSSxi}}

\figure{$f^{(\chi)}_1/f^{(\chi)}_0$ for various values of $\beta$.
The continuus line shows the $\xi\rightarrow\infty$ extrapolation.
\label{f1chi}}

\figure{$f^{(\xi)}_1/f^{(\xi)}_0$ for various values of $\beta$.
The continuus line shows the $\xi\rightarrow\infty$ extrapolation.
\label{f1xi}}

\figure{$f^{(P)}_0$ versus $R/L$ at $\beta=0.86$.
\label{fP0}}

\figure{$g^{(P)}_0$ versus $R/L$ at $\beta=0.86$.
\label{gP0}}

\figure{$\phi(L/\xi_0)$  at $\beta=0.86$.
\label{phi}}

\figure{$\psi(R/L)$  at $\beta=0.86$ for several values of $L$.
\label{psi}}

\figure{$\Delta_{(\lambda)}^{-1}(k,0;L)$ versus $k\xi_0$ for several values
of $z\equiv L/\xi_0$.
\label{deltak}}

\figure{$\Delta_{(\lambda)}^{-1}(0;L)$ versus $L/\xi_0$.
The dashed line shows the function $(L/\xi_0)^{-2}$.
\label{delta0}}

\figure{$\phi_Y(L/\xi_0)$  for $\xi_0=16.04$.
\label{phiY}}

\figure{$\psi_Y(R/L)$  for $\xi_0=16.04$ and for several values of $L$.
\label{psiY}}

\figure{The large-$N$ string tension on the lattice
at finite $\xi_0$. $N\sigma_0\xi_0^2$ is plotted versus $1/\xi_0^2$.
\label{sigma0}}

\figure{Finite size scaling of the magnetic susceptibility for $N=100$.
The dashed and the continuus lines show respectively
the leading and the next-to-leading order of the large-$N$ expansion.
\label{fchi100}}

\figure{Finite size scaling of the correlation length for $N=100$.
The dashed and the continuus lines show respectively
the leading and the next-to-leading order of the large-$N$ expansion.
\label{fxi100}}

\figure{Finite size scaling of the magnetic susceptibility for $N=41$.
The dashed and the continuus lines show respectively
the leading and the next-to-leading order of the large-$N$ expansion.
\label{fchi41}}

\figure{Finite size scaling of the correlation length for $N=41$.
The dashed and the continuus lines show respectively
the leading and the next-to-leading order of the large-$N$ expansion.
\label{fxi41}}

\figure{Finite size scaling of the magnetic susceptibility for $N=21$.
The dashed and the continuus lines show respectively
the leading and the next-to-leading order of the large-$N$ expansion.
\label{fchi21}}

\figure{Finite size scaling of the correlation length for $N=21$.
The dashed and the continuus lines show respectively
the leading and the next-to-leading order of the large-$N$ expansion.
\label{fxi21}}

\figure{Finite size scaling of the magnetic susceptibility for $N=10$.
The dashed and the continuus lines show respectively
the leading and the next-to-leading order of the large-$N$ expansion.
\label{fchi10}}

\figure{Finite size scaling of the correlation length for $N=10$.
The dashed and the continuous lines show respectively
the leading and the next-to-leading order of the large-$N$ expansion.
\label{fxi10}}

\figure{Finite size scaling function
of the magnetic susceptibility obtained by
Monte Carlo simulations for several values of $N$. The dashed line
shows the leading order of the large-$N$ expansion.
\label{fchi}}

\figure{Finite size scaling function of the correlation length obtained by
Monte Carlo simulations for several values of $N$. The dashed line
shows the leading order of the large-$N$ expansion.
\label{fxi}}

\figure{Finite size scaling function of the magnetic susceptibility
versus $L/(\xi N^{1/3})$.
\label{fchivrs}}

\figure{Finite size scaling function of the correlation length
versus $L/(\xi N^{1/3})$.
\label{fxivrs}}

\figure{Rescaled Creutz ratios versus $r=R/\xi$ for $N=100$.
Data were taken at $\beta=0.54$ and on a $L=48$ lattice.
The continuous and the dashed
lines show the large-$N$ predictions obtained at $\beta=0.54$
respectively for an infinite and a $L=48$ lattice.
The dotted line is the $\beta\rightarrow\infty$
prediction, with asymptotic value $\pi/N$.
\label{sigma100}}


\mediumtext
\begin{table}
\caption{Large-$N$ string tension versus $\beta$.}
\label{tab_sigma}
\begin{tabular}{r@{}lr@{}lr@{}l}
\multicolumn{2}{c}{$\beta$}&
\multicolumn{2}{c}{$\xi_0$}&
\multicolumn{2}{c}{$N\sigma_0\xi_0^2$}\\
\tableline
 0.&54   &  2.&1474   & 3.&2709  \\
 0.&65   &  4.&2859   & 3.&1827 \\
 0.&70   &  5.&8678   & 3.&1658 \\
 0.&75   &  8.&0337   & 3.&1557  \\
 0.&81   & 11.&7123   & 3.&1489  \\
 0.&86   & 16.&0350   & 3.&1457  \\
 0.&97   & 32.&0067   & 3.&14284  \\
 &$\infty$& &$\infty$ & 3.&14159...\\
\end{tabular}
\end{table}

\narrowtext

\narrowtext

\begin{table}
\caption{Summary of the simulation results for the ${\rm CP}^9$ model.}
\label{tab_cp9}
\begin{tabular}{r@{}lrr@{}lr@{}lr@{}l}
\multicolumn{2}{c}{$\beta$}&
\multicolumn{1}{r}{$L$}&
\multicolumn{2}{c}{$E$}&
\multicolumn{2}{c}{$\xi$}&
\multicolumn{2}{c}{$\chi_{\rm m}$}\\
\tableline
0&.75 & 18  & 0&.7157(3)  & 3&.71(2)   &  19&.39(11) \\
 &    & 21  & 0&.7179(2)  & 3&.61(2)   &  18&.66(10) \\
 &    & 24  & 0&.7196(2)  & 3&.43(2)   &  17&.62(8) \\
 &    & 27  & 0&.7198(2)  & 3&.34(2)   &  17&.16(8) \\
 &    & 30  & 0&.7202(1)  & 3&.289(15) &  16&.94(5) \\
 &    & 42  & 0&.7202(1)  & 3&.27(3)   &  16&.80(5) \\
 &    & 48  & 0&.7202(1)  & 3&.25(3)   &  16&.77(4) \\
 &    & 54  & 0&.7202(1)  & 3&.31(3)   &  16&.80(4) \\
 &    & 60  & 0&.7203(1)  & 3&.31(4)   &  16&.79(4) \\
\tableline
0&.8 &  9  & 0&.6473(5)  & 3&.377(9)  &  17&.25(4)\\
 &   & 12  & 0&.6539(4)  & 4&.070(13) &  22&.88(7)\\
 &   & 18  & 0&.6606(3)  & 5&.02(2)   &  30&.88(12)\\
 &   & 24  & 0&.6642(2)  & 5&.19(3)   &  32&.19(23)\\
 &   & 27  & 0&.6650(2)  & 5&.18(3)   &  32&.14(21)\\
 &   & 30  & 0&.6659(2)  & 5&.07(4)   &  31&.07(23)\\
 &   & 33  & 0&.6668(2)  & 4&.79(3)   &  29&.29(17)\\
 &   & 36  & 0&.6667(1)  & 4&.70(4)   &  28&.75(17)\\
 &   & 39  & 0&.6669(1)  & 4&.67(3)   &  28&.56(13)\\
 &   & 48  & 0&.66689(6) & 4&.60(2)   &  28&.14(8) \\
 &   & 60  & 0&.66699(4) & 4&.60(2)   &  28&.08(5) \\
\tableline
0&.85 & 15 & 0&.6145(3)   & 5&.221(16) &  34&.20(10) \\
 &    & 36 & 0&.6213(1)   & 7&.21(5)   &  53&.63(43) \\
 &    & 48 & 0&.62196(9)  & 6&.63(5)   &  48&.73(32)\\
 &    & 60 & 0&.62216(8)  & 6&.48(5)   &  47&.41(26)\\
 &    & 72 & 0&.62214(8)  & 6&.44(6)   &  47&.18(22) \\
\end{tabular}
\end{table}

\narrowtext

\begin{table}
\caption{Summary of the simulation results for the ${\rm CP}^{20}$ model.}
\label{tab_cp20}
\begin{tabular}{r@{}lrr@{}lr@{}lr@{}l}
\multicolumn{2}{c}{$\beta$}&
\multicolumn{1}{r}{$L$}&
\multicolumn{2}{c}{$E$}&
\multicolumn{2}{c}{$\xi$}&
\multicolumn{2}{c}{$\chi_{\rm m}$}\\
\tableline
0&.65 &  6  & 0&.7550(4)  &  2&.041(3)   &  8&.562(13) \\
 &    &  9  & 0&.7767(4)  &  2&.611(6)   & 11&.96(3) \\
 &    & 12  & 0&.7864(3)  &  2&.989(8)   & 13&.99(4) \\
 &    & 15  & 0&.7914(2)  &  3&.193(8)   & 14&.84(4) \\
 &    & 18  & 0&.7950(2)  &  3&.186(17)  & 14&.49(8) \\
 &    & 21  & 0&.7976(2)  &  3&.003(19)  & 13&.43(8) \\
 &    & 24  & 0&.7990(2)  &  2&.835(17)  & 12&.67(6) \\
 &    & 30  & 0&.7992(1)  &  2&.721(13)  & 12&.26(4) \\
 &    & 36  & 0&.7993(1)  &  2&.696(13)  & 12&.17(3) \\
 &    & 42  & 0&.7995(1)  &  2&.691(18)  & 12&.14(3) \\
\tableline
0&.7 &  6 & 0&.7004(6)   &  2&.239(4)  &  9&.767(17) \\
 &   &  9 & 0&.7191(4)   &  2&.957(7)  & 14&.65(3) \\
 &   & 12 & 0&.7267(3)   &  3&.524(7)  & 18&.58(4) \\
 &   & 30 & 0&.7388(2)   &  4&.19(6)   & 21&.74(25) \\
 &   & 36 & 0&.7389(2)   &  3&.83(3)   & 19&.9(1) \\
 &   & 48 & 0&.7392(2)   &  3&.716(25) & 19&.51(8) \\
\end{tabular}
\end{table}

\narrowtext

\begin{table}
\caption{Summary of the simulation results for the ${\rm CP}^{40}$ model.}
\label{tab_cp40}
\begin{tabular}{r@{}lrr@{}lr@{}lr@{}l}
\multicolumn{2}{c}{$\beta$}&
\multicolumn{1}{r}{$L$}&
\multicolumn{2}{c}{$E$}&
\multicolumn{2}{c}{$\xi$}&
\multicolumn{2}{c}{$\chi_{\rm m}$}\\
\tableline
0&.57 &  6  & 0&.8440(6)   & 1&.764(3)   &    7&.049(12) \\
 &    &  9  & 0&.8672(3)   & 2&.163(4)   &    8&.876(19) \\
 &    & 12  & 0&.8781(2)   & 2&.358(4)   &    9&.44(2) \\
 &    & 15  & 0&.8830(2)   & 2&.406(8)   &    9&.30(3) \\
 &    & 18  & 0&.8863(2)   & 2&.292(7)   &    8&.67(2) \\
 &    & 21  & 0&.8888(2)   & 2&.078(9)   &    7&.97(2) \\
 &    & 24  & 0&.8887(1)   & 2&.032(8)   &    7&.85(15) \\
 &    & 33  & 0&.8889(1)   & 2&.004(7)   &    7&.755(9) \\
 &    & 42  & 0&.8890(1)   & 2&.011(10)  &    7&.756(8) \\
\tableline
0&.6  &  6  & 0&.8007(6)  &  1&.897(4)   &   7&.853(16) \\
 &    &  9  & 0&.8235(5)  &  2&.398(6)   &  10&.51(3) \\
 &    & 12  & 0&.8340(3)  &  2&.701(6)   &  11&.84(3) \\
 &    & 15  & 0&.8385(2)  &  2&.864(7)   &  12&.25(3) \\
 &    & 18  & 0&.8415(2)  &  2&.902(10)  &  12&.07(4) \\
 &    & 42  & 0&.8454(1)  &  2&.431(11)  &  10&.105(15) \\
\end{tabular}
\end{table}

\narrowtext

\begin{table}
\caption{Summary of the simulation results for the ${\rm CP}^{99}$ model.}
\label{tab_cp99}
\begin{tabular}{r@{}lrr@{}lr@{}lr@{}l}
\multicolumn{2}{c}{$\beta$}&
\multicolumn{1}{r}{$L$}&
\multicolumn{2}{c}{$E$}&
\multicolumn{2}{c}{$\xi$}&
\multicolumn{2}{c}{$\chi_{\rm m}$}\\
\tableline
0&.54 &  6  & 0&.8749(4)  &  1&.670(2)   &    6&.590(9) \\
 &    &  9  & 0&.8988(2)  &  2&.037(4)   &    8&.11(2) \\
 &    & 12  & 0&.9088(2)  &  2&.221(5)   &    8&.53(2) \\
 &    & 15  & 0&.9135(2)  &  2&.263(5)   &    8&.344(16) \\
 &    & 18  & 0&.9164(1)  &  2&.210(6)   &    7&.936(17) \\
 &    & 21  & 0&.9177(1)  &  2&.111(5)   &    7&.541(12) \\
 &    & 24  & 0&.9188(2)  &  2&.011(7)   &    7&.278(13) \\
 &    & 36  & 0&.91899(7) &  1&.916(8)   &    7&.078(7) \\
 &    & 42  & 0&.91895(5) &  1&.916(8)   &    7&.072(5) \\
 &    & 48  & 0&.91893(7) &  1&.922(9)   &    7&.069(6) \\
\end{tabular}
\end{table}

\end{document}